\documentclass[preprint]{elsarticle}

\usepackage{lineno,hyperref}
\usepackage{amsmath}
\usepackage{tikz}
\modulolinenumbers[5]

\journal{\href{https://arxiv.org}{arXiv.org}}

\begin{document}

\begin{frontmatter}

\title{Two Dimensional Edge Detection by Guided Mode Resonant Metasurface}

\author[sharif]{Amirhossein Saba}
\author[sharif]{Mohammad Reza Tavakol}
\author[sharif]{Parisa Karimi-Khoozani}
\author[sharif]{Amin Khavasi\corref{cor1}}
\ead{khavasi@sharif.edu}
\cortext[cor1]{Corresponding author}
\address[sharif]{Department of Electrical Engineering, Sharif University of Technology, P.O. Box 11555-4363, Tehran, Iran}

\begin{abstract}
In this letter, a new approach to perform edge detection is presented using an all-dielectric CMOS-compatible metasurface. The design is based on guided-mode resonance which provides a high quality factor resonance to make the edge detection experimentally realizable. The proposed structure that is easy to fabricate, can be exploited for detection of edges in two dimensions due to its symmetry. Also, the trade-off between gain and resolution of edge detection is discussed which can be adjusted by appropriate design parameters. The proposed edge detector has also the potential to be used in ultrafast analog computing and image processing.
\end{abstract}


\end{frontmatter}


Recently, researchers have succeeded in overcoming relatively large size and slow response of electronic and mechanical analog computers by introducing a new concept of analog optical computing. This concept has been used to all-optical performing of different real-time computations and operations in the temporal or spatial domain. A plethora of works have been done on temporal analog optical computing including optical temporal differentiator based on ring resonator \cite{liu2008compact}, photonic temporal integrator \cite{slavik2008photonic} and all-optical differential equation solver \cite{yang2014all}. On the other side, for spatial computation, the two approaches proposed in \cite{silva2014performing} allow implementation of the desired transfer function in the spatial domain using metasurfaces or metamaterials with engineered meta-atoms or a multilayered slab that is transversely homogenous but longitudinally inhomogeneous. Furthermore, the spatial differentiation, as one of the primary operators for optical processing has presented in \cite{doskolovich2014spatial} using phase-shifted Bragg grating and in \cite{youssefi2016analog} based on Brewster effect at oblique incidence.

Recently, Zhu et al. have proposed a plasmonic-based spatial differentiator and have shown that it can be used for edge detection, a method in which intensity changes in an image is described \cite{torre1986edge}, without any Fourier lens \cite{zhu2017plasmonic}.  Edge detection, as the first step in object detection, simplifies the image processing by decreasing the under processing data \cite{canny1986computational}.  However, time-consuming computation of high-throughput edge detection requirement represents a key challenge in applications with real-time image processing \cite{zhu2017plasmonic}. The differentiator proposed by Zhu et al. \cite{zhu2017plasmonic} has a high gain because of  a narrow plasmonic resonance which is due to the excitation of a surface plasmon.  The high gain is a very important feature which makes the edge detection experimentally viable by overcoming inevitable noise. However, their design requires fine-tuning of geometrical and material parameters because it works based on critical coupling condition, i.e. the equality of radiative leakage rate of the structure and the intrinsic loss rate of the plasmonic material. Moreover, their structure detects edges merely in one dimension, owing to the fact that it’s based on the one-dimensional first-order derivative of the incident beam profile \cite{zhu2017plasmonic}. To realize two-dimensional edge detection, one can use the two-dimensional differentiators like Laplace operator. For example, this operator has been synthesized by phase-shift Bragg grating in \cite{bykov2014optical}, but for increasing the gain of the structure, several numbers of Bragg layers is necessary that increase the size and fabrication complexity of the structure.

In this paper, we propose an all-dielectric CMOS-compatible ultrathin metasurface with two-dimensional periodicity to realize two-dimensional edge detection. Our structure is in light of guided mode resonance (GMR) \cite{peng1996resonant,liu1998high} and its fabrication is easy.  The high quality factor GMR is exploited for providing high-gain edge detection. The amount of gain that affects the bandwidth of edge detection can be controlled arbitrarily by adjusting design parameters of the structure for the desired wavelength.

\begin{figure}[t]
\centering
\includegraphics[width=0.9\linewidth]{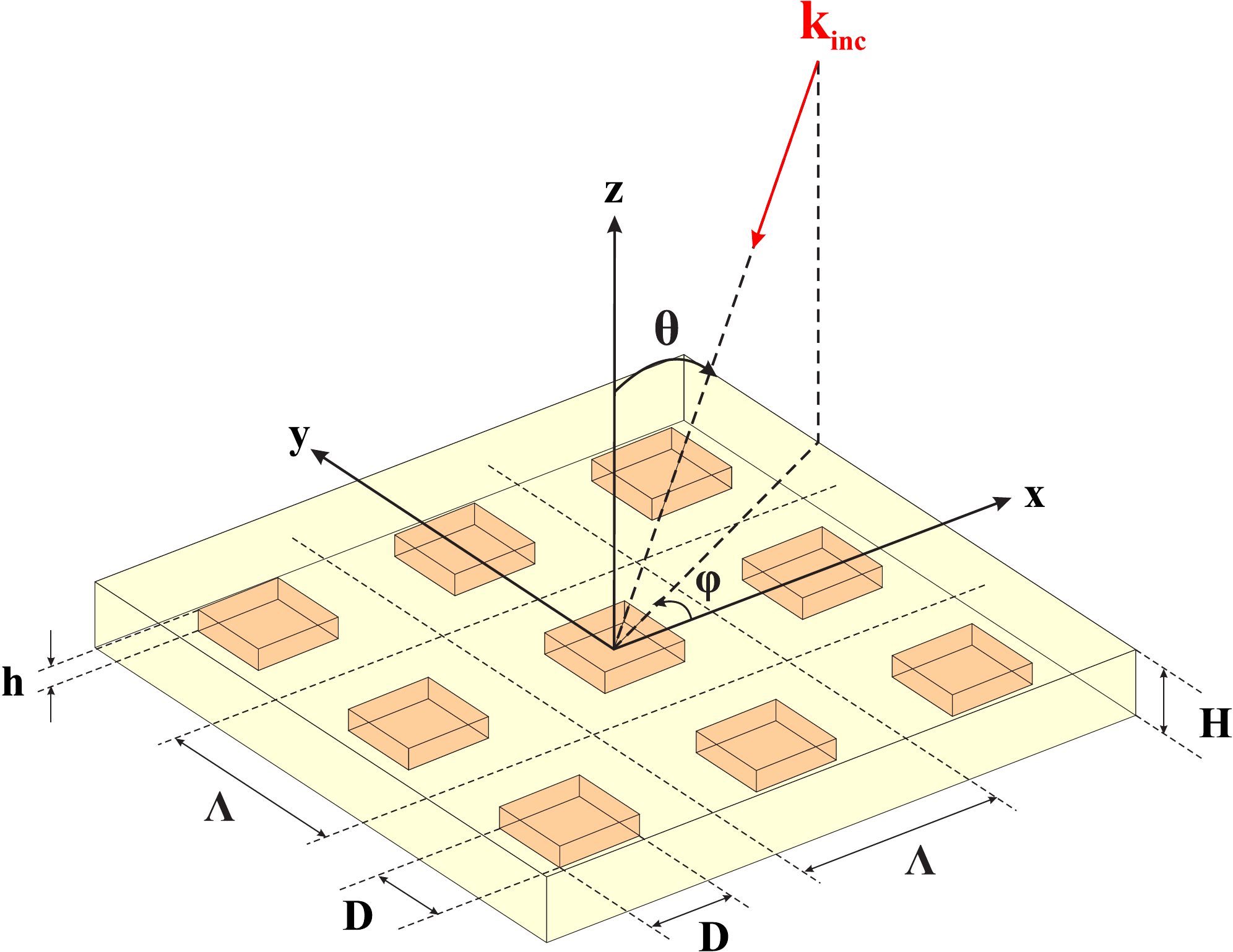}\llap{\makebox[22cm][c]{\raisebox{9cm}{{\textbf{(a)}}}}}
\\[8mm]
\includegraphics[width=0.9\linewidth]{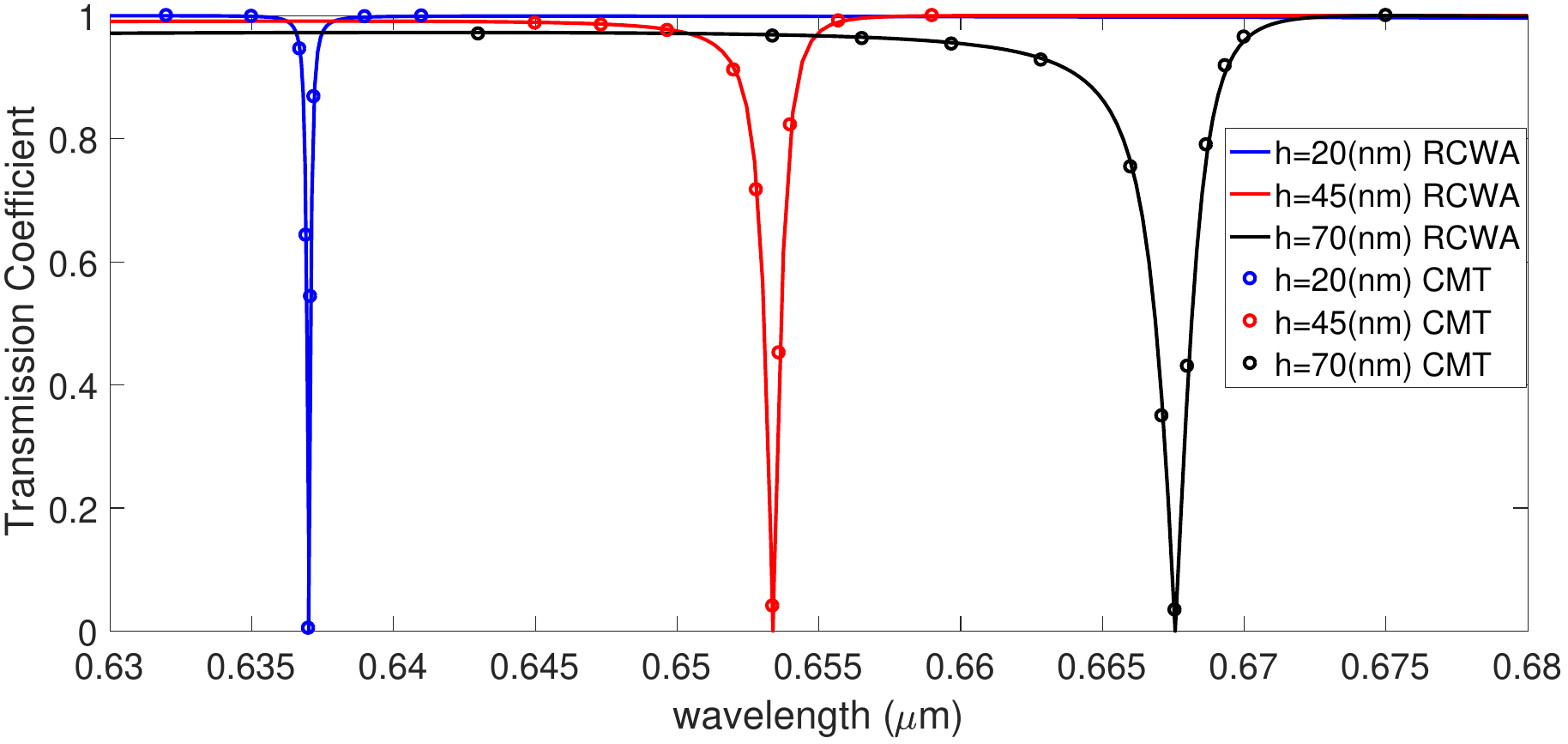}\llap{\makebox[22cm][c]{\raisebox{6cm}{{\textbf{(b)}}}}}
\caption{(a) Schematic of metasurface illuminated by a plane wave. (b) Zeroth-order transmission coefficient for $h=20\ {\rm{nm}}$ (blue), $h=45\ {\rm{nm}}$ (red) and $h=70\ {\rm{nm}}$ (black) calculated by RCWA (solid) and temporal CMT (circles) at normal incidence. Other parameters are assumed as $\Lambda  = 500\ {\rm{nm}}$, $D = 0.5\Lambda $ and $H = h + 200\ {\rm{nm}}$.}
\label{fig:Fig1}
\end{figure}

Fig.~\hyperref[fig:Fig1]{1(a)} illustrates the metasurface consisting of a slab of silicon dioxide (${\rm{SiO}}_2$) with height $H$ and the refractive index of $n=1.46$ that functions as the cover and substrate for periodic silicon nitride inclusions. These inclusions with the refractive index of $n=2$, height $h$, and width $D$ are arranged symmetrically at the center of silicon dioxide slab with identical periods of $\Lambda$ in both $x$ and $y$ directions.

The GMR condition is satisfied when the incident plane wave couples to a leaky waveguide mode due to the periodicity of the structure. This condition for the normal incidence can be formulated as,
\begin{equation}
\sqrt {{{(m{k_\Lambda })}^2} + {{(n{k_\Lambda })}^2}}  = {\beta _r}
\label{eq:ref1}
\end{equation}
in which, m and n are integers, ${k_\Lambda } = \frac{{2\pi }}{\Lambda }$ is the grating wave-vector, and $\beta_r$ is the real part of the propagation constant of the leaky mode which depends on the wavelength because of the modal dispersion. 

By appropriate selection of period, one can observe resonance under the normal incidence, leading to a dip in the transmission coefficient of the metasurface. The first resonance is depicted in the Fig.~\hyperref[fig:Fig1]{1(b)} for 3 different values of $h$ assuming $\Lambda  = 500\ {\rm{nm}}$, $D = 0.5\Lambda $ and $H = h + 200\ {\rm{nm}}$. These results have been calculated by rigorous coupled wave analysis (RCWA) \cite{hugonin2005reticolo} for TE-polarized normal incidence. Due to the x-y symmetry of the structure, the transmission coefficient will be the same for TM polarization at normal incidence. As it is obvious from Fig.~\hyperref[fig:Fig1]{1(b)}, the height of the inclusions can be increased to achieve less leaky waveguide mode that leads to a broader linewidth and lower quality factor. Not only does the height of the inclusions affect the linewidth, but also it changes the central wavelength of the resonance due to its influence on the propagation constant in Eq.~(\ref{eq:ref1}). Nevertheless, one can adjust the period to achieve the resonance in a given wavelength. For this purpose, we can calculate the resonant wavelength by Eq.~(\ref{eq:ref1}) in which the dispersion of propagation constant can be approximately obtained using effective medium theory \cite{bertoni1989frequency}.

In addition to numerical analysis, the transmission coefficient, depicted as circles in Fig.~\hyperref[fig:Fig1]{1(b)}, has been calculated by phenomenological analysis of temporal coupled mode theory (CMT) \cite{fan2003temporal} which predicts the shape of the transmission coefficient with a good agreement with the RCWA results.
\begin{figure}[t]
\centering
\includegraphics[width=0.9\linewidth]{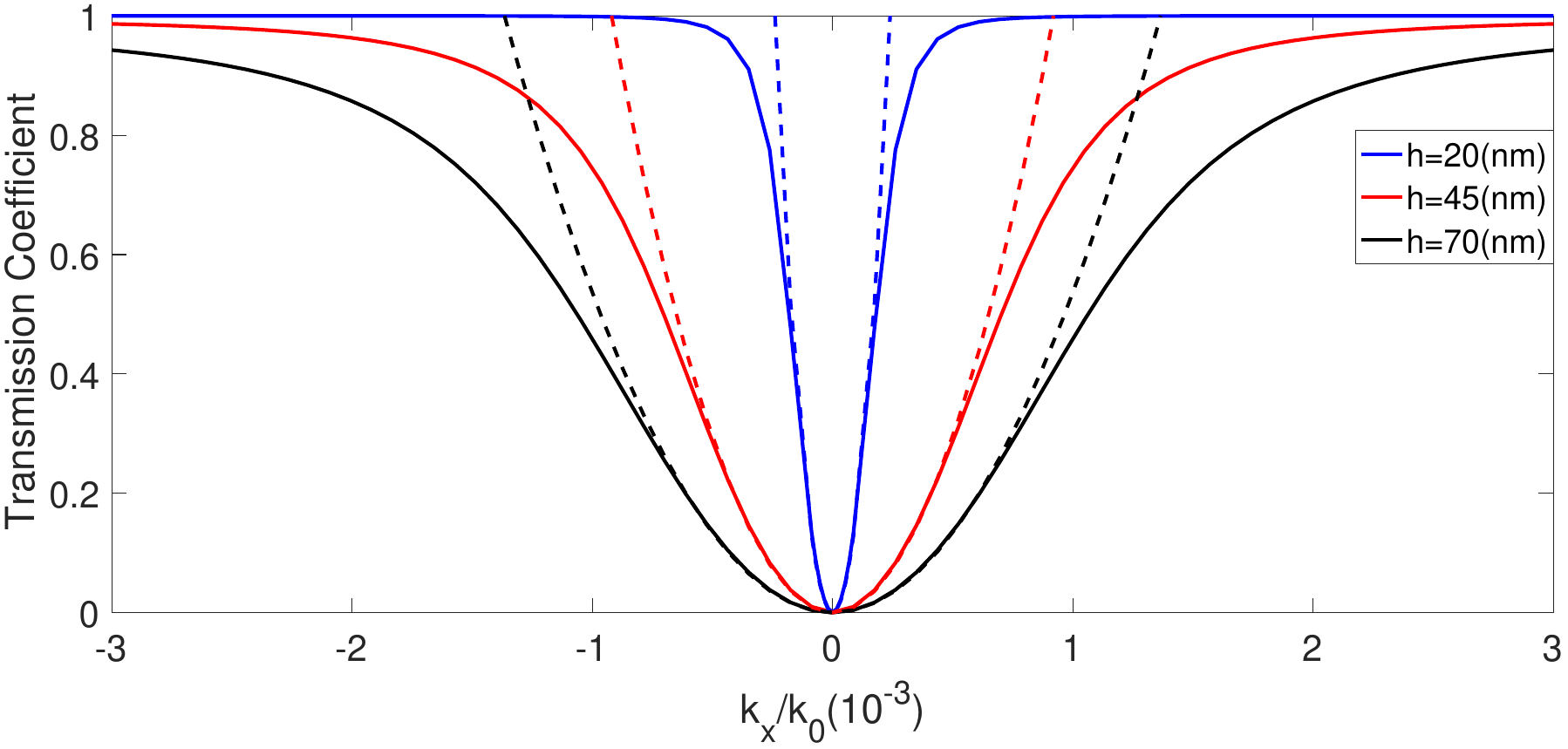}
\caption{Spatial transfer function spectra (solid) in one dimension calculated by RCWA and the parabolic fitting (dashed) for $\varphi = 0^\circ$. The parameters of the metasurface are $\Lambda = 500\ {\rm{nm}}$, $D=0.5\Lambda$, $H=h+200\ {\rm{nm}}$ and $h=20\ {\rm{nm}}$ at $\lambda = 637\ {\rm{nm}}$, $h=45\ {\rm{nm}}$ at $\lambda = 653\ {\rm{nm}}$, $h=70\ {\rm{nm}}$ at $\lambda = 668\ {\rm{nm}}$. The incident beam is TE polarized.}
\label{fig:Fig2}
\end{figure}

Now, we consider a normal incident beam illuminating the structure. We can decompose this beam into TE and TM-polarized beams and express them as the superposition of plane waves by spatial Fourier transform,
\begin{equation}
\begin{aligned}
{{\vec E}^{\rm{inc}}}(x,y,z) =& \vec E_{\rm{TE}}^{\rm{inc}}(x,y,z) + \vec E_{\rm{TM}}^{\rm{inc}}(x,y,z) \\
=&\iint{{{{\tilde E}_{\rm{TE}}}({k_x},{k_y}){{\vec u}_{\rm{TE}}}({k_x},{k_y}){e^{i\vec k.\vec r}}d{k_x}d{k_y}}} \\
&+\iint{{{{\tilde E}_{\rm{TM}}}({k_x},{k_y}){{\vec u}_{\rm{TM}}}({k_x},{k_y}){e^{i\vec k.\vec r}}d{k_x}d{k_y}}}
\end{aligned}
\label{eq:ref2}
\end{equation}
in which, $\vec k = {k_x}\hat x + {k_y}\hat y + {k_z}\hat z$  is the wave-vector, ${\vec u_{\rm{TM}}} =  - \cos \varphi \cos \theta \hat x - sin\varphi \cos \theta \hat y + sin\theta \hat z$ and ${\vec u_{\rm{TE}}} =  - \sin \varphi \hat x + \cos \varphi \hat y$ are the unitary vectors associated with TM and TE polarizations in which $\theta$ and $\varphi$ are elevation and azimuthal angles specifying the angle of incidence according to Fig.~\hyperref[fig:Fig1]{1(a)}. In addition, ${\tilde E_{\rm{TE}}}({k_x},{k_y})$ and ${\tilde E_{\rm{TM}}}({k_x},{k_y})$ are spatial frequency spectra of TE and TM-polarized beams that can be calculated given the polarization of the incident beam and its spatial spectra \cite{bykov2014optical}. If we assume that the incident beam is linearly polarized such that all of its plane wave components are TE-polarized, we can express it in Cartesian coordinate system as follows,
\begin{equation}
\begin{aligned}
{{\vec E}^{\rm{inc}}}(x,y,z) &= E_x^{\rm{inc}}(x,y,z)\hat x + E_y^{\rm{inc}}(x,y,z)\hat y \\
&=\iint{{[{{\tilde E}_x}({k_x},{k_y})\hat x + {{\tilde E}_y}({k_x},{k_y})\hat y]{e^{i\vec k.\vec r}}d{k_x}d{k_y}}}
\end{aligned}
\label{eq:ref3}
\end{equation}
where
\begin{subequations}
\begin{align}
{{\tilde E}_x}({k_x},{k_y}) &= \frac{{ - {k_y}}}{{\sqrt {k_x^2 + k_y^2} }}{{\tilde E}_{\rm{TE}}}({k_x},{k_y})\\
{{\tilde E}_y}({k_x},{k_y}) &= \frac{{{k_x}}}{{\sqrt {k_x^2 + k_y^2} }}{{\tilde E}_{\rm{TE}}}({k_x},{k_y})
\end{align}
\label{eq:ref4}
\end{subequations}
Owing to the fact that the incident beam has only TE-polarized plane wave components, the transmitted beam is simplified as,
\begin{equation}
\begin{aligned}
&{{\vec E}^{\rm{trans}}}(x,y,z) = \\
&\iint{[{{\tilde E}_{\rm{TE}}}({k_x},{k_y}){{\vec u}_{\rm{TE}}}({k_x},{k_y})]{T_{\rm{TE}}}({k_x},{k_y}){e^{i\vec k.\vec r}}d{k_x}d{k_y}}= \\
&\iint{[{{\tilde E}_x}({k_x},{k_y})\hat x + {{\tilde E}_y}({k_x},{k_y})\hat y]{T_{\rm{TE}}}({k_x},{k_y}){e^{i\vec k.\vec r}}d{k_x}d{k_y}}
\end{aligned}
\label{eq:ref5}
\end{equation}
where ${T_{\rm{TE}}}({k_x},{k_y})$ is the transmission coefficient for the TE-polarized incident plane wave which its incident angle is specified by $k_x$ and $k_y$.
\begin{figure}[t]
\centering
\includegraphics[width=0.85\linewidth]{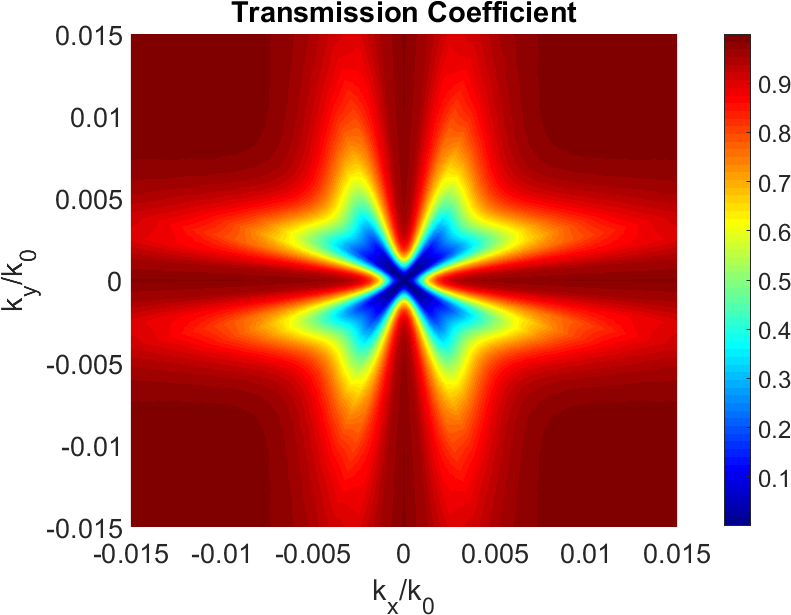}\llap{\makebox[22cm][c]{\raisebox{8cm}{{\textbf{(a)}}}}}
\\[5mm]
\includegraphics[width=0.85\linewidth]{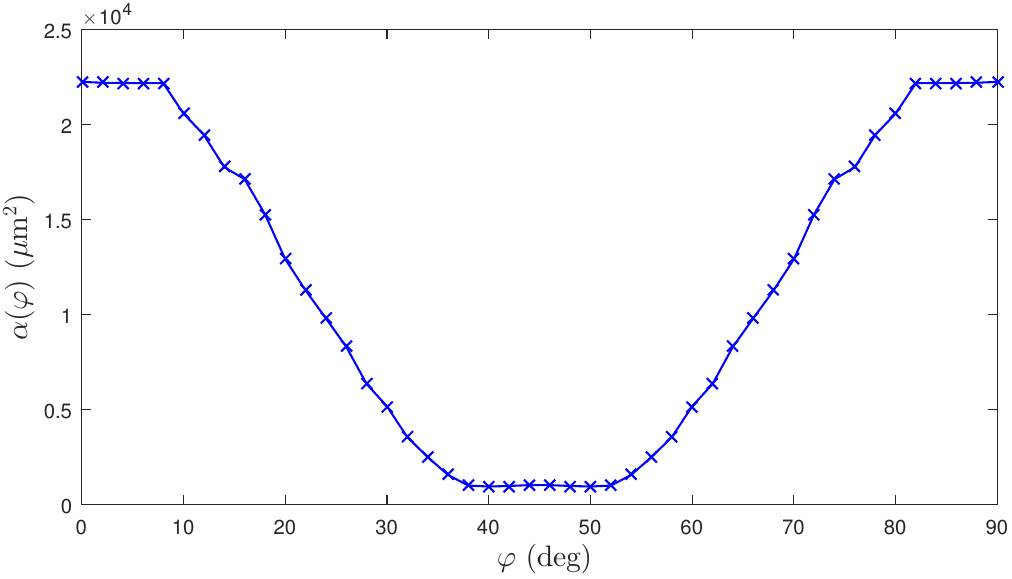}\llap{\makebox[22cm][c]{\raisebox{6cm}{{\textbf{(b)}}}}}
\caption{(a) Two dimensional spatial transfer function spectra and (b) gain of edge detection versus $\varphi$. Results are calculated by RCWA for $h=70\ {\rm{nm}}$ at $\lambda = 668\ {\rm{nm}}$ and assuming  $\Lambda = 500\ {\rm{nm}}$, $D=0.5\Lambda$ and $H = h+200\ {\rm{nm}}$. The incident beam is TE polarized.}
\label{fig:Fig3}
\end{figure}

Therefore, for an incident beam with described spatial Fourier spectra given by Eq.~(\ref{eq:ref4}), the transmitted beam is polarized in the same direction and the Fourier spectra of its Cartesian components can be expressed as,
\begin{subequations}
\begin{align}
\tilde E_x^{\rm{trans}}({k_x},{k_y}) &= {T_{\rm{TE}}}({k_x},{k_y})\tilde E_x^{\rm{inc}}({k_x},{k_y})\\
\tilde E_y^{\rm{trans}}({k_x},{k_y}) &= {T_{\rm{TE}}}({k_x},{k_y})\tilde E_y^{\rm{inc}}({k_x},{k_y})
\end{align}
\label{eq:ref6}
\end{subequations}

Based on RCWA results, $T_{\rm{TE}}$ has been depicted for 3 values of $h$ as a function of ${k_x}/{k_0}$ for $\varphi = 0^\circ$ in Fig.~\ref{fig:Fig2} (the other parameters are $\Lambda  = 500\ {\rm{nm}}$, $D = 0.5\Lambda $ and $H = h + 200\ {\rm{nm}}$). Because $T_{\rm{TE}}$ is zero for ${k_x} = {k_y} = 0$ and due to symmetries, it can be expressed as a parabolic curve of $k_x$ (for $k_y=0$) retaining the first non-zero term in Taylor expansion. This curve fitting is accurate within a specific bandwidth. As it obvious in Fig.~\ref{fig:Fig2} the spatial bandwidth of the resonance, as similar to the temporal bandwidth, can be controlled by the appropriate height of inclusions. The significance of bandwidth control will be further elucidated when the role of the resonance bandwidth in the edge detection is discussed.

Fig.~\ref{fig:Fig3} illustrates $T_{\rm{TE}}$ versus ${k_x}/{k_0}$ and ${k_y}/{k_0}$ for $\Lambda  = 500\ {\rm{nm}}$, $D = 0.5\Lambda $, $h=70\ {\rm{nm}}$ and $H=270\ {\rm{nm}}$ at $\lambda=668\ {\rm{nm}}$ (the resonant wavelength). According to this figure we can consider ${T_{{\rm{TE}}}}({k_x},{k_y}) = {T_{{\rm{TE}}}}({k_y},{k_x})$ as well as ${T_{{\rm{TE}}}}({k_x},{k_y}) = {T_{{\rm{TE}}}}( - {k_y},{k_x}) = {T_{{\rm{TE}}}}( - {k_x}, - {k_y}) = {T_{{\rm{TE}}}}({k_y}, - {k_x})$ that are due to symmetries of the structure. It is simpler to express ${T_{{\rm{TE}}}}$ as a function of the two parameters of polar coordinate i.e. $\varphi  = {\tan ^{ - 1}}({k_y}/{k_x})$ ($0 \le \varphi  < 180$) and ${k_r} =  \pm \sqrt {k_x^2 + k_y^2} $ instead of $k_x$ and $k_y$. Due to the fact that ${T_{{\rm{TE}}}}(\varphi ,{k_r} = 0)$ is zero at the resonance and based on the symmetries, the first non-zero polynomial approximation of ${T_{{\rm{TE}}}}(\varphi ,{k_r})$ is,
\begin{equation}
{T_{{\rm{TE}}}}(\varphi ,{k_r}) = \alpha (\varphi )k_r^2 = \alpha (\varphi )\left( {k_x^2 + k_y^2} \right)
\label{eq:ref7}
\end{equation}

Based on the Eq.~(\ref{eq:ref6}) and the form of Eq.~(\ref{eq:ref7}), we can obtain the edge-detected image as the output of the metasurface and define $\alpha (\varphi )$ as the gain of the edge detection. We plotted $\alpha (\varphi )$ in Fig.~\hyperref[fig:Fig3]{3(b)} which shows lower gain around $\varphi  = {45^\circ }$. 
\begin{figure}[t]
\centering
\includegraphics[width=0.45\linewidth]{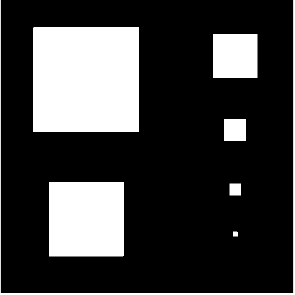}\llap{\makebox[11.8cm][c]{\raisebox{5.5cm}{{\textbf{(a)}}}}}\llap{\makebox[9mm][c]{\raisebox{1mm}{\tikz \fill [white] (0.0,0.0) rectangle (0.7,0.18);}}} 
\llap{\makebox[5.2cm][c]{\raisebox{3.8cm}{\color{white}{\textbf{1}}}}}\llap{\makebox[5.75cm][c]{\raisebox{1.2cm}{\color{white}{\textbf{2}}}}}\llap{\makebox[1.2cm][c]{\raisebox{4.25cm}{\color{white}{\textbf{3}}}}}\llap{\makebox[1.2cm][c]{\raisebox{2.9cm}{\color{white}{\textbf{4}}}}}\llap{\makebox[1.2cm][c]{\raisebox{1.8cm}{\color{white}{\textbf{5}}}}}\llap{\makebox[1.2cm][c]{\raisebox{0.95cm}{\color{white}{\textbf{6}}}}}
\hfill
\includegraphics[width=0.45\linewidth]{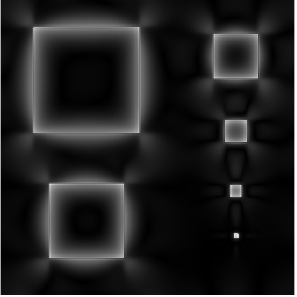}\llap{\makebox[11.8cm][c]{\raisebox{5.5cm}{{\textbf{(b)}}}}}\llap{\makebox[9mm][c]{\raisebox{1mm}{\tikz \fill [white] (0.0,0.0) rectangle (0.7,0.18);}}}
\\[2mm]
\begin{tabular}{|c|c|c|c|c|c|}
	\hline
    1     & 2     & 3     & 4     & 5     & 6 \\ \hline
    765   & 525   & 320   & 150   & 75    & 35 \\ \hline
\end{tabular}
\\[5mm]
\includegraphics[width=0.45\linewidth]{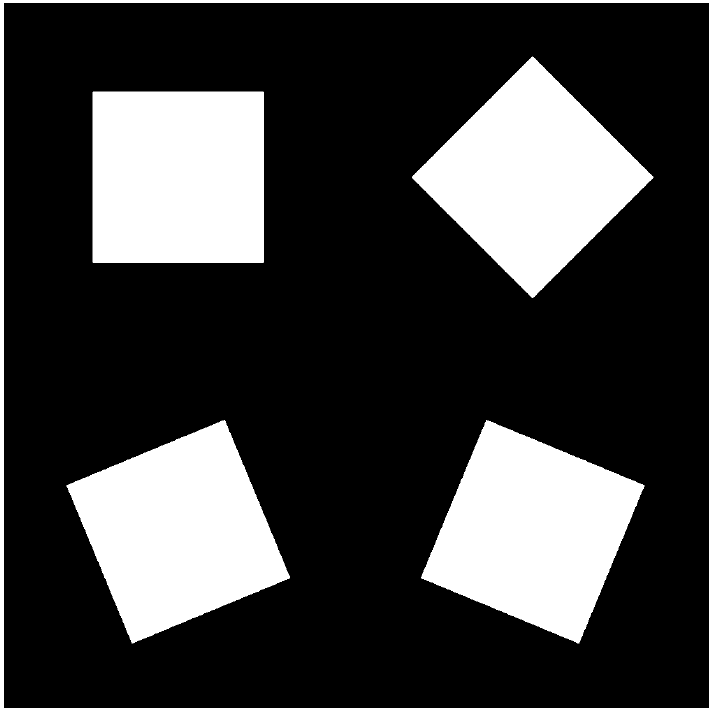}\llap{\makebox[11.8cm][c]{\raisebox{5.5cm}{{\textbf{(c)}}}}}\llap{\makebox[9mm][c]{\raisebox{1mm}{\tikz \fill [white] (0.0,0.0) rectangle (0.7,0.18);}}} 
\llap{\makebox[8.4cm][c]{\raisebox{3.95cm}{\color{black}{\textbf{1}}}}}\llap{\makebox[3.0cm][c]{\raisebox{3.95cm}{\color{black}{\textbf{3}}}}}\llap{\makebox[8.4cm][c]{\raisebox{1.2cm}{\color{black}{\textbf{2}}}}}\llap{\makebox[3.0cm][c]{\raisebox{1.2cm}{\color{black}{\textbf{4}}}}}
\hfill
\includegraphics[width=0.45\linewidth]{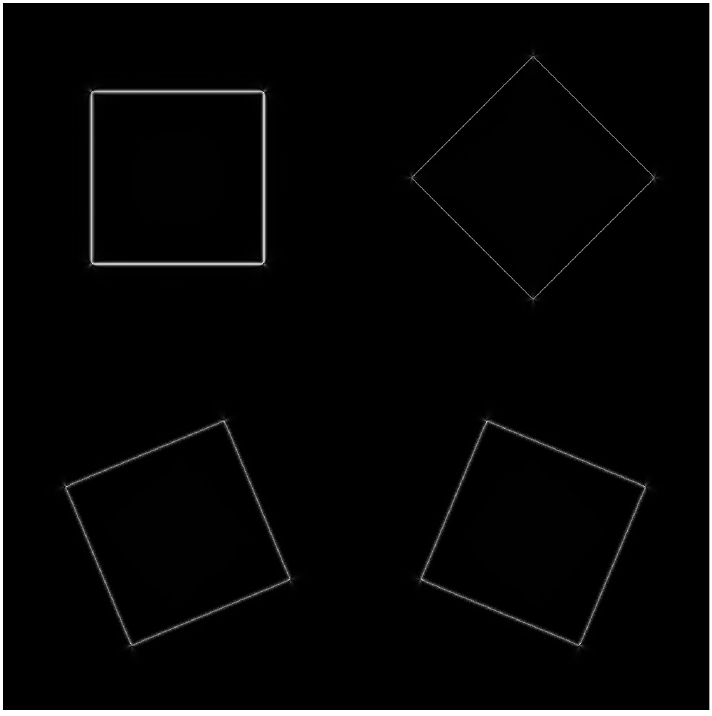}\llap{\makebox[11.8cm][c]{\raisebox{5.5cm}{{\textbf{(d)}}}}}\llap{\makebox[9mm][c]{\raisebox{1mm}{\tikz \fill [white] (0.0,0.0) rectangle (0.7,0.18);}}}
\\[2mm]
\begin{tabular}{|c|c|c|c|}
	\hline
    1     & 2     & 3     & 4 \\ \hline
    $0^\circ$     & $22.5^\circ$  & $45^\circ$    & $67.5^\circ$ \\ \hline
\end{tabular}
\caption{(a) Square-shaped input beams with different sizes and (b) their corresponding outputs. The inset table: size of the square-shaped input beams in $\mu$m. (c) Rotated input patterns and (d) their corresponding outputs showing fader edges as the rotation angle approaches to $\varphi = 45^\circ$ . The white bar shows the length of 270 $\mu$m in (a) and (b) and the length of 6.4 mm in (c) and (d). The inset table: rotation angles of the square-shaped input beams in degree.}
\label{fig:Fig4}
\end{figure}

In Fig.~\hyperref[fig:Fig4]{4(a)} we depicted square-shaped input beams with different sizes and calculated the transmitted images by transfer function of Fig.~\ref{fig:Fig3} corresponding to these beams in Fig.~\hyperref[fig:Fig4]{4(b)}. It can be seen that the edges in both dimensions are resolved for large enough beams. However, the edges of the smaller beams are not well resolved. Smaller beams have a higher spatial frequency content being out of the device bandwidth in spatial frequency domain.  So there is a trade-off between the resolution and the gain.

For silicon nitride patches with height $h=70$ nm, the minimum size of the beams of Fig.~\hyperref[fig:Fig4]{4(a)} that can be resolved is approximately 335 $\mu$m in light of Rayleigh criterion. It should be noted that by changing the height of the inclusions, the bandwidth of the resonance can be controlled that leads to a change in this resolution.

\begin{figure}[t]
\centering
\includegraphics[width=0.85\linewidth]{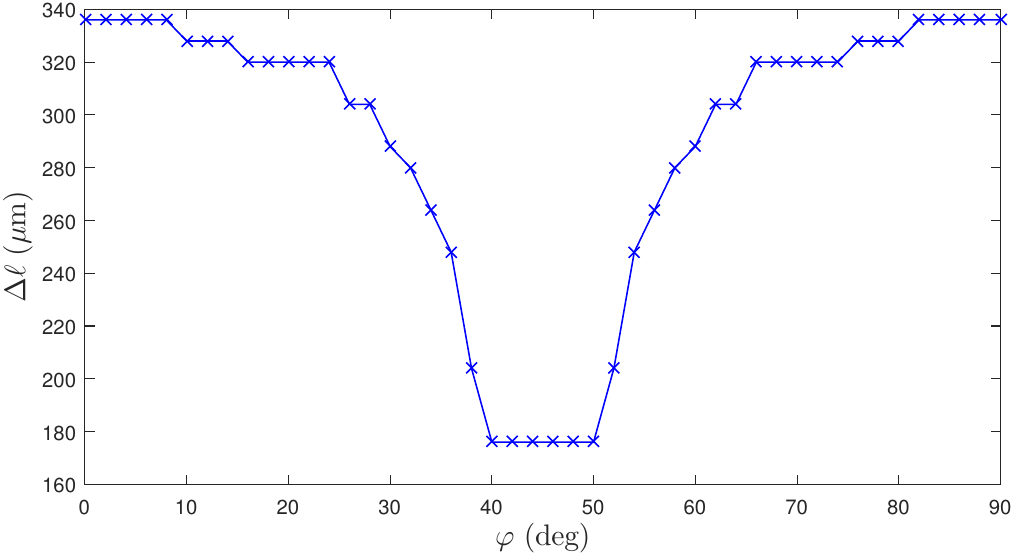}
\caption{Rayleigh spatial resolution versus $\varphi$ for transfer function of Fig.~\hyperref[fig:Fig3]{3(a)}}
\label{fig:Fig5}
\end{figure}

It is worth mentioning that according to Figs.~\hyperref[fig:Fig3]{3 (a), (b)} for $\varphi=45^\circ$ the metasurface provides edge detection with minimum gain but with maximum bandwidth. For highlighting this effect, we rotate the incident beams for $22.5^\circ$, $45^\circ$ and $67.5^\circ$ in Fig.~\hyperref[fig:Fig4]{4(c)}. As expected, the transmitted images in Fig.~\hyperref[fig:Fig4]{4(d)} represent lower intensity (darker edges) as the rotation angle approaches to $45^\circ$. However, the lower gain leads to a higher bandwidth and as a result to a better resolution which is plotted as a function of $\varphi$ in Fig.~\ref{fig:Fig5}. According to this figure and as our expectation, the resolution is better around $\varphi = 45^\circ$.
\begin{figure}[t]
\centering
\includegraphics[width=0.45\linewidth]{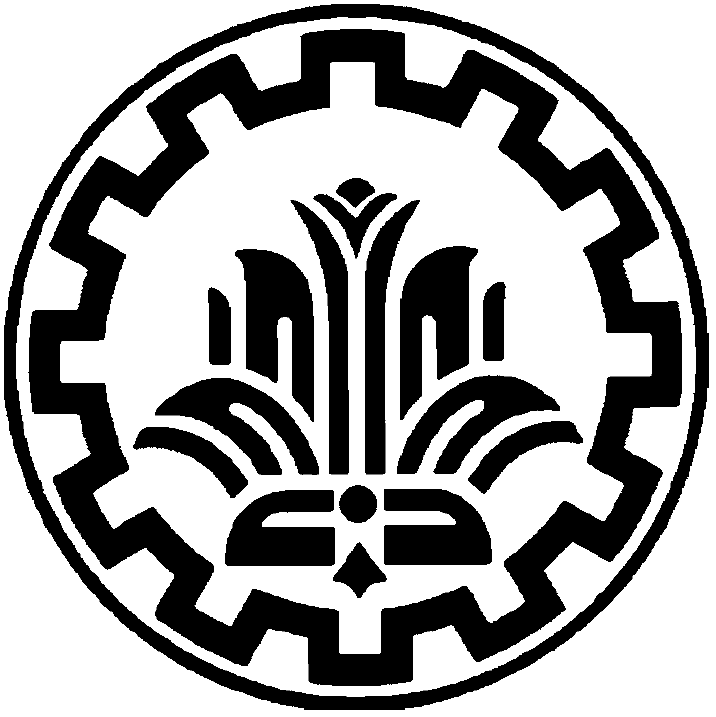}\llap{\makebox[11.5cm][c]{\raisebox{5.6cm}{{\textbf{(a)}}}}}\llap{\makebox[9mm][c]{\raisebox{1mm}{\tikz \fill [black] (0.0,0.0) rectangle (0.7,0.18);}}} 
\hfill
\includegraphics[width=0.45\linewidth]{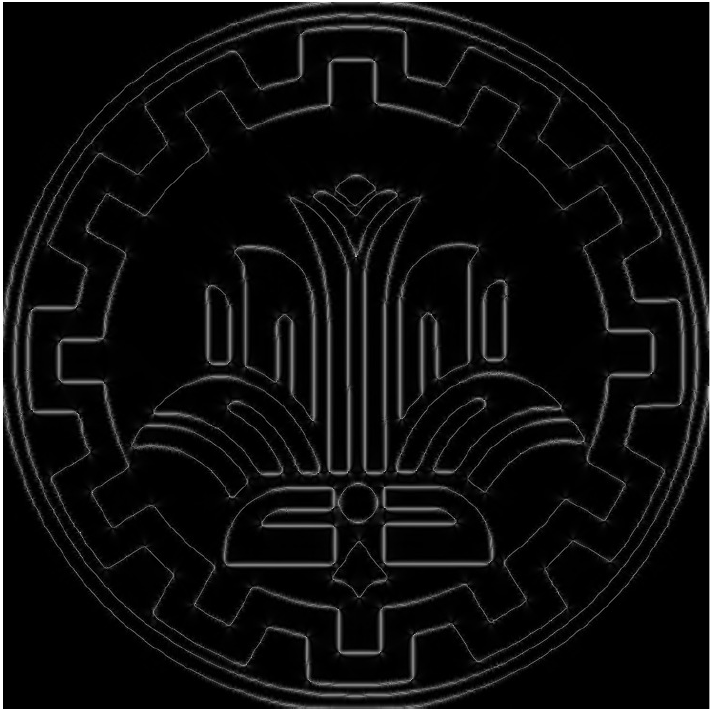}\llap{\makebox[11.5cm][c]{\raisebox{5.6cm}{{\textbf{(b)}}}}}\llap{\makebox[9mm][c]{\raisebox{1mm}{\tikz \fill [white] (0.0,0.0) rectangle (0.7,0.18);}}}
\caption{(a) Sharif University of Technology logo as the input incident beam. (b) Edges of the image resolved in the transmitted beam. The back and white bars correspond to the length of 4.8 mm.}
\label{fig:Fig6}
\end{figure}

Finally, we evaluate edge detection using proposed metasurface on the Sharif University of Technology logo as illustrated in Fig.~\hyperref[fig:Fig6]{6(a)}. The transmitted image amplitude in Fig.~\hyperref[fig:Fig6]{6(b)} demonstrates edges of the incident image. Although the transmitted image is bereft of exactly uniform amplitude because of gain dropping around $\varphi=45^\circ$, the more quality of edge detection around this angle is clear from the transmitted image which is due to the higher bandwidth of device around this angle.

In summary, we proposed a CMOS-compatible metasurface that can be used for edge detection with high-gain. This high-gain edge detection makes the detection experimentally feasible even for images with low-gradient edges and in the presence of unavoidable noise.  The proposed structure can be used for edge detection in two dimension. Although the intensity of edges is not uniform in different directions, the information about existence of edges of an image in all directions can be provided by considering a threshold amplitude in the transmitted image. In addition, with respect to the size of the objects in the input images, the guided mode resonance with appropriate bandwidth can be selected based on the gain-resolution trade-off.


\bibliography{mybibfile}

\end{document}